\documentclass[prl,aps,epsfig,twocolumn,showpacs]{revtex4}
\usepackage{graphicx}
\usepackage{amssymb}
\usepackage{amsmath}

\begin{document}

\title{Anomalous magneto-optical Kerr effect in perpendicularly magnetized Co/Pt films on
two-dimensional colloidal crystals}
\author{Z. Liu, L. Shi, Z. Shi, X. H. Liu, J. Zi$^{*}$, and S. M. Zhou$^{*}$\footnotetext{${}^{*}${Electronic
mail:jzi@fudan.edu.cn; shimingzhou@yahoo.com}}}
\address{Surface Physics Laboratory (National Key Laboratory) and Department~of~Physics, Fudan University,
Shanghai 200433, China}

\author{S. J. Wei and J. Li}
\address{Department of Optical Science and Engineering, Fudan University, Shanghai 200433, China }

\date{\today}

\vspace{5 cm}

\begin{abstract}

\indent We have studied the magneto-optical Kerr effect and
optical reflectance of perpendicular magnetized Co/Pt films on
 self-assembly two-dimensional polystyrene spheres. It is shown that
the magneto-optical and the reflectance spectra are correlated to
each other. Calculations have shown that the magneto-optical Kerr
rotation and the Kerr ellipticity spectra are governed by the
resonant coupling of light to the excitations of surface plasmon
polaritons and modified by the sphere diameters. These effects
lead to the possibility of developing new magnetoplasmonic
nanosensors.

\end{abstract}

\vspace{5 cm}


\maketitle
\begin{figure}[tb]
\begin{center}
\resizebox*{2 in}{!}{\includegraphics*{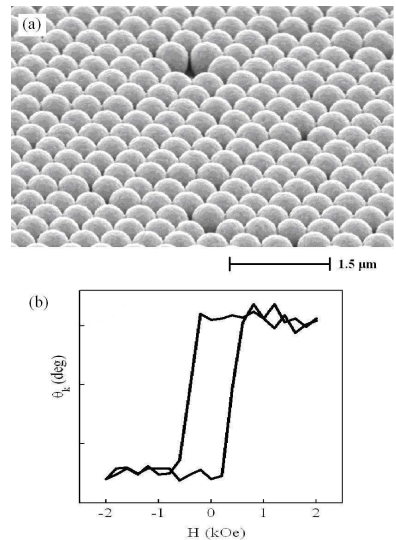}} \caption{SEM
image of the self-assembly array of spheres on a plain quartz
chip, where $d=500$ nm, (a). Typical polar Kerr rotation loop of
the nanostructured [Co/Pt]$_{\mathrm{3}}$ sample. The wavelength
is 800 nm, (b). } \label{fig1}
\end{center}
\end{figure}
\begin{figure}[tb]
\begin{center}
\resizebox*{2 in }{!}{\includegraphics*{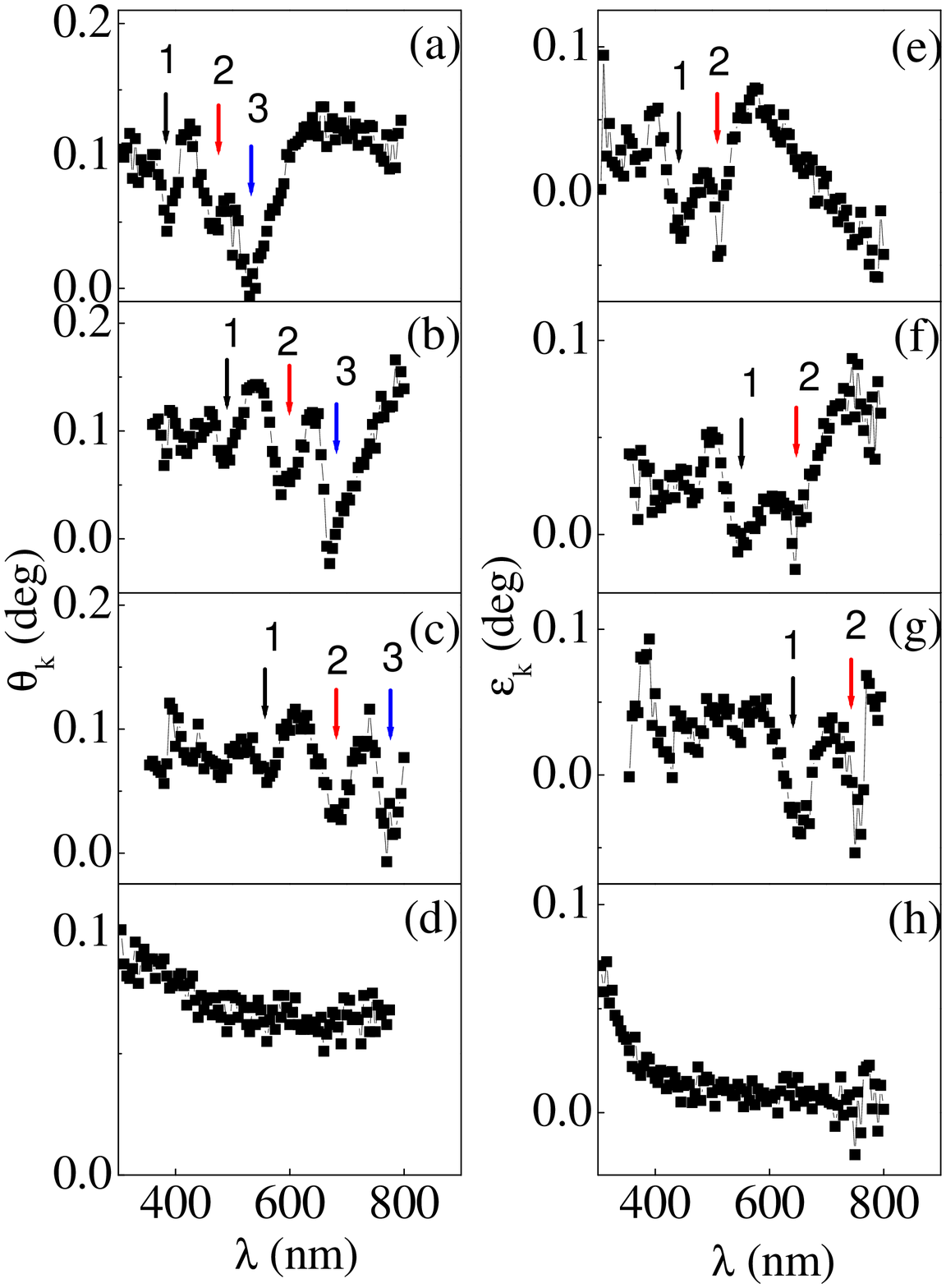}}
\end{center}
\caption{$\theta_{\mathrm{K}}$ (left column) and
$\varepsilon_{\mathrm{K}}$ (right column) spectra for the
nanostructured [Co/Pt]$_{\mathrm{3}}$ films with $d=400$ nm (a,
e), 500 nm (b, f), and 600 nm (c, g). For comparison,
$\theta_{\mathrm{K}}$ (d) and $\varepsilon_{\mathrm{K}}$ (h)
spectra of the Co/Pt continuous films are also given.}
\label{fig2}
\end{figure}
\begin{figure}[tb]
\begin{center}
\resizebox*{2 in}{!}{\includegraphics*{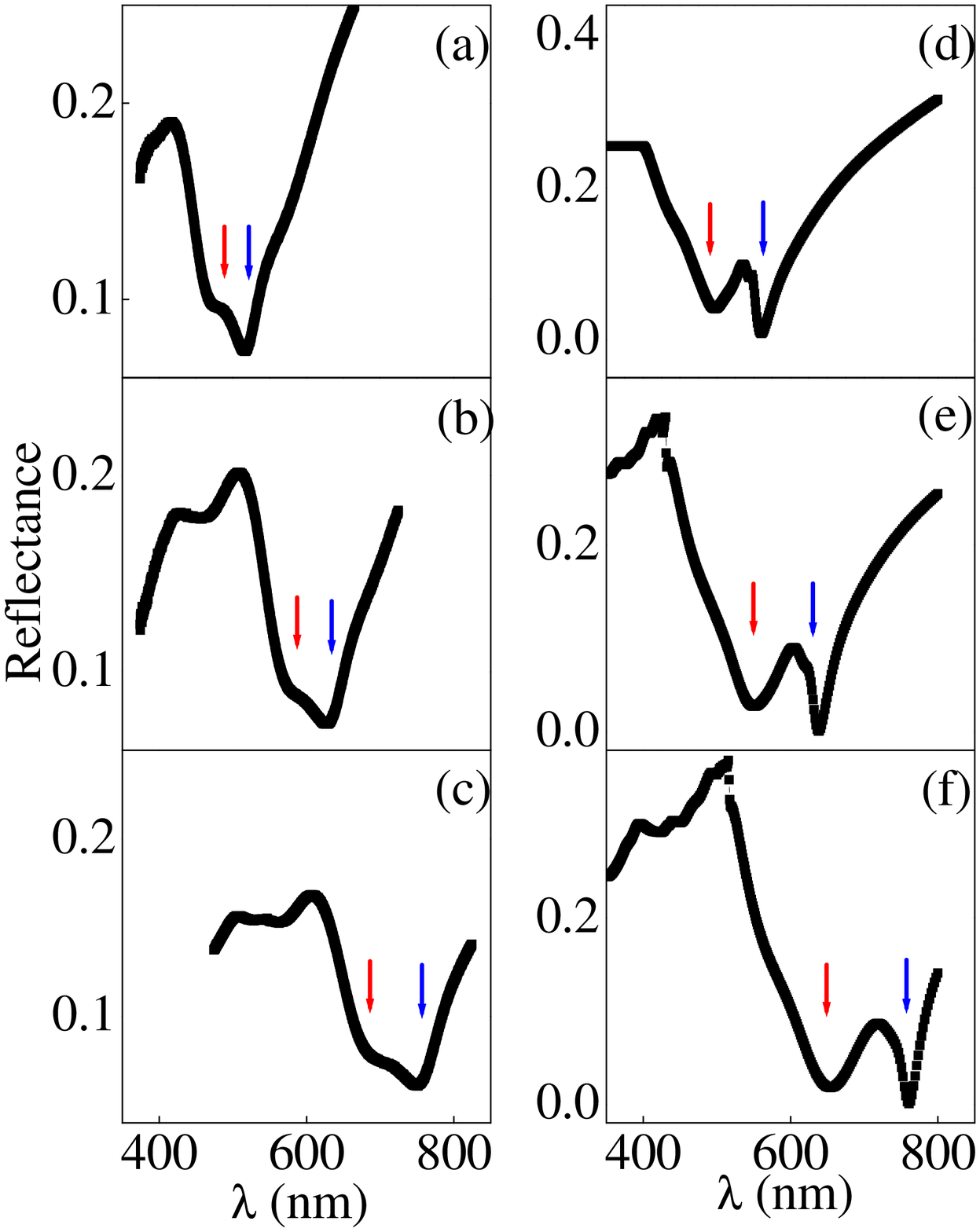}}
\end{center}
\caption{Measured (left column) and calculated (right column)
optical reflectance spectra of the nanostructured
[Co/Pt]$_{\mathrm{3}}$ films with $d=400$ nm (a, d), 500 nm (b,
e), and 600 nm (c, f).} \label{fig3}
\end{figure}

\indent As one kind of the key functional composite materials,
periodic nanostructures have attracted much interest since
extraordinary optical transmission of light was first reported in
an optically opaque metallic film perforated with a
two-dimensional (2D) array of subwavelenth holes by Ebbesen et
al~\cite{Ebbesen1998,Lezec2002,Barnes2003,Lomakin2005}. This
phenomenon is generally attributed to the coupling of surface
plasmon polaritons (SPPs) to the light. The optical response of
periodic structures can be easily tailored because it strongly
depends on the lattice symmetry, the metallic film thickness, and
the refractive index of the dielectric media. Hence, these
nanostructures are of crucial importance in both the underlying
physics and the perceived potential applications in nanophotonics,
quantum-information processing, nanolithography, and
surface-enhanced Raman
scattering~\cite{Altewischer2002,Srituravanich2004,Brolo2004,Zhan2006}.\\
\indent Very recently, the effect of the SPPs on magneto-optics
has also attracted much
attention~\cite{Zvezdin1997,Strelniker1999,Diwekar2004,Inoue2006,Gonzalez-diaz2008}.
For self-assembly arrays of Au/Co/Au
nanosandwiches~\cite{Gonzalez-diaz2008}, as high sensitivity
magneto-plasmonic biosensors, the magneto-optical Kerr effect
(MOKE) is governed by the surface plasmonic resonance. It has been
predicted theoretically that the MOKE can be enhanced greatly due
to the surface plasmonic resonance. However, the MOKE of these
plasmonic systems is usually smaller than corresponding continuous
magnetic films. For Co films perforated with a subwavelength hole
array, for example, the MOKE is much smaller than that of uniform
Co films~\cite{Diwekar2004}. At the same time, the large MOKE is
essential in practical applications. Therefore, it is required to
further study the effect of SPPs on the MOKE. Moreover, the
perpendicular magnetic anisotropy is the prerequisite for
applications of the magnetoplasmonic systems. Up to data, however,
most of these magnetic nanostructures only have in-plane magnetic
anisotropy. In this Letter, Co/Pt multilayers were deposited on
self-assembly array of polystyrene sphere prepared by nanosphere
lithography. It is noted that the Co/Pt continuous films with thin
Co layers may have perpendicular magnetic anisotropy and
reasonably large MOKE~\cite{Carcia1989,Zeper1989}. The new
structure is induced in the MOKE spectra by the surface plasmonic
resonance, compared with
corresponding continuous films.\\
\indent The nanostructured Co/Pt films~\cite{Love2002,Liu2005}
were prepared in two steps. First, the 2D polystyrene sphere
arrays were prepared by injecting an aqueous solution of colloidal
dispersion with a suitable concentration into a channel that was
formed by two parallel quartz slides separated by a U-shaped
spacer, where the monodisperse polystyrene microspheres (size
dispersion 1\%) used here were purchased from Duke Scientific
Corps. The quartz slides were pretreated to render their surface
hydrophilic by soaking in a solution of 30\% hydrogen peroxide at
$80^{o}C$ for 30 min. After drying in air, highly ordered colloid
crystals were grown within the channel under capillary force.
Secondly, Pt(10 nm)(bottom)/[Co(1 nm)/Pt(2 nm)]$_{\mathrm{3}}$
multilayers were then deposited by dc magnetron sputtering at
ambient temperature on the arrays of the polystyrene microspheres.
In the sputtering system, the base pressure of the system was
$2\times 10^{-5}$ Pa and the Ar pressure 0.4 Pa during deposition.
The microbeads were hemispherically covered with the Co/Pt films
and close-packed array of half-shells was formed, where the sphere
diameter ($d$) can be conveniently controlled from 200 nm to
several micrometers. In comparison, the same Co/Pt multilayers
were also deposited on plain quartz substrates. Sample structures
were characterized by scanning electron microscopy (SEM). The
polar Kerr rotation $\theta_{\mathrm{K}}$ and Kerr ellipticity
$\varepsilon_{\mathrm{K}}$ were recorded by a magneto-optical Kerr
spectrometer with an applied field of 10 kOe perpendicular to the
film plane~\cite{Cheng1999}. Reflectance spectra were measured by
using a microspectrophotometer~\cite{Zi2003}. In all reflection
measurements, the optical spot size on the samples was about 1.0
mm. All measurements were performed at room temperature.\\
\indent Figure~\ref{fig1} (a) shows the SEM image of the
nanostructured Co/Pt sample. The order of the sphere arrays is
clearly demonstrated. In the upper part of the image, there is a
vacancy through which the monolayer colloid crystal substrate can
be clearly identified. Since the polystyrene microspheres were
densely packed, the Co/Pt film half-shells on adjacent spheres
were interconnected after a certain amount of deposition, thus
forming a conduct in network on the spheres. Due to the locally
curved surface of the spheres, a lateral variation of the metal
thickness was created on the spheres, with the thinnest layer at
the equator of each sphere. Figure~\ref{fig1} (b) shows typical
polar Kerr loop of the nanostructured Co/Pt sample. Apparently,
the sample shows
perpendicular magnetic anisotropy with the out-of-plane coercivity of 400 Oe.\\
\indent The $\theta_{\mathrm{K}}$ and $\varepsilon_{\mathrm{K}}$
spectra of the nanostructured Co/Pt films are shown in
Fig.~\ref{fig2}. In comparison, the results of the Co/Pt
continuous films are also given. Few distinguished features can be
found. Firstly, for the continuous films, $\theta_{\mathrm{K}}$
and $\varepsilon_{\mathrm{K}}$ change monotonically with the
wavelength. On the contrary, for the nanostructured Co/Pt samples,
$\theta_{\mathrm{K}}$ and $\varepsilon_{\mathrm{K}}$ both undergo
several maxima and minima with increasing wavelength. Secondly,
although the spectra of $\theta_{\mathrm{K}}$ and
$\varepsilon_{\mathrm{K}}$ are similar for samples with different
$d$, either $\theta_{\mathrm{K}}$ or $\varepsilon_{\mathrm{K}}$
minima are shifted in translation towards long wavelengths with
increasing $d$. The $\theta_{\mathrm{K}}$ minima are located at
383 nm, 464 nm, and 540 nm for $d=400$ nm; 480 nm, 580 nm, and 668
nm for $d=500$ nm; and 556 nm, 684 nm, and 771 nm for $d=600$ nm.
The $\varepsilon_{\mathrm{K}}$ minima are located at 442 nm and
513 nm for d=400 nm; 543 nm and 645 nm for d=500 nm; 656 nm and
758 nm for d=600 nm. Apparently, the anomalous MOKE spectra are
induced by the 2D sphere arrays and modified by varying $d$.
Finally, at long wavelengths, the Kerr rotation of the
nanostructured Co/Pt multilayers with small $d$ is enhanced, in
comparison with that
Co/Pt continuous films. \\
\indent In order to interpret above anomalous MOKE results, the
optical reflectance spectra of the nanostructured Co/Pt samples
were measured in the zero-order reflection at normal incidence,
where the reflectance was normalized to that of a pure quartz
substrate. Figures~\ref{fig3}(a)-~\ref{fig3}(c) show typical
experimental results of the nanostructures Co/Pt samples with
different $d$. The reflectance has a minimum at 513 nm, 624 nm,
and 747 nm for $d=400$ nm, 500 nm, and 600 nm, respectively. It is
noted that the reflectance changes monotonically with the
wavelength for the continuous Co/Pt films. Apparently, the
anomalous reflectance spectra are induced by the 2D colloid
crystal structure.

\begin{figure}[tb]
\begin{center}
\resizebox*{2 in}{!}{\includegraphics*{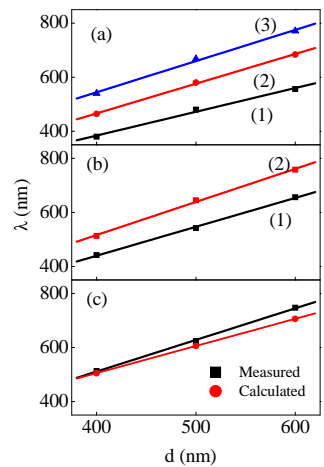}}
\end{center}
\caption{The shift of the $\theta_{\mathrm{K}}$(a),
$\varepsilon_{\mathrm{K}}$(b), and reflectance (c) minima with
$d$. The inset numbers in (a) and (b) refer to the minimum marks
in Figs.2$\&$ 3.} \label{fig4}
\end{figure}

\indent To get deep insight into the mechanism of the anomalous
reflectance spectra, we have calculated the reflectance spectra of
2D colloid crystals with different lattice constant (400 nm-600
nm) by a finite-difference time-domain method~\cite{Taflove1995},
as shown in Figs.~\ref{fig3}(d)-~\ref{fig3}(f). The refractive
index of the polystyrene spheres is assumed to be 1.59 in the
entire wavelength region studied here and the wavelength
dependence of the refractive index of the Co/Pt multilayers on
plain quartz substrate was measured by
ellipsometry~\cite{Chen1994}. The calculated reflectance spectra
reproduce the main features of the experimental results.
Importantly, the predicted minimum positions are in good agreement
with the experimental observations.\\
\indent It is significant to compare the red-shift of the minimal
$\theta_{\mathrm{K}}$, $\varepsilon_{\mathrm{K}}$, and reflectance
for the nanostructured Co/Pt samples. Figure~\ref{fig4} shows that
the minima of $\theta_{\mathrm{K}}$, $\varepsilon_{\mathrm{K}}$,
and reflectance are almost shifted in the same trend. Firstly, as
the fingerprint of the SPPs, the red-shift of the structure in the
reflectance spectra is shifted with increasing $d$. This is
because the periodic structure of the metallic layer serves to
induce excitations of the SPPs, thereby leading to the
extraordinary optical transmittance, that is to say, due to
surface plasmonic resonance coupling,  the wavelength location of
the reflectance minimum is expected to be proportional to $d$.
With the incomplete shell morphology, the present microstructures
allow strong coupling of the SPPs with the colloid crystal, which
is affected by the dielectric properties of the colloidal spheres.
Therefore, the optical response of ordered array of dielectric
spheres partially coated with Co/Pt film is a superposition of
scattering diffraction and light reradiation via the excitations
of SPPs on the metallic films. Secondly, the structure observed in
the MOKE spectra, which is caused by the thin magnetic layer, are
correlated to that of the reflectance spectra as a function of
$d$~\cite{Gonzalez-diaz2008}, as shown in Fig.~\ref{fig4}. It is
noted that the minima of $\theta_{\mathrm{K}}$,
$\varepsilon_{\mathrm{K}}$, and the reflectance are located at
different wavelengths. This is because the refractive index and
the magneto-optical constants of Co/Pt films have different
wavelength dependence. The strong decrement of the polar Kerr
effect in the spectral range near the plasmonic resonance band is
caused by the interplay of the magneto-optical activities and the
resonant coupling of light with the SPPs in the nanostructured
Co/Pt films. \\
\indent In conclusion, we have measured the spectra of the polar
Kerr rotation, the Kerr ellipticity, and the reflectance in the
visible region for the nanostructured Co/Pt samples with
perpendicular magnetic anisotropy. The MOKE and the reflectance of
the nanostructured Co/Pt samples undergo several minima with
wavelength, which are shifted in translation with $d$. Theoretical
simulations show that the structure of the MOKE and the optical
reflectance spectra is caused by the resonant coupling of light to
surface plasmonic excitations. This work may
facilitate developing new magnetoplasmonic nanosensors with multiplexing capabilities.\\











\end{document}